# Rocky Worlds Limited to ~1.8 Earth Radii by Atmospheric Escape During a Star's Extreme UV Saturation

Owen R. Lehmer[1,2], David C. Catling[1]
(1) Dept. Earth and Space Sciences, Box 351310, University of Washington, Seattle, WA
(2) MS 239-4, Space Science Division, NASA Ames Research Center, Moffett Field, CA

Recent observations and analysis of low mass (<10 $M_\oplus$), exoplanets have found that rocky planets only have radii up to 1.5-2 $R_\oplus$. Two general hypotheses exist for the cause of the dichotomy between rocky and gas-enveloped planets (or possible water worlds): either low mass planets do not necessarily form thick atmospheres of a few wt. %, or the thick atmospheres on these planets easily escape driven by x-ray and extreme ultraviolet (XUV) emissions from young parent stars. Here we show that a cutoff between rocky and gas-enveloped planets due to hydrodynamic escape is most likely to occur at a mean radius of 1.76±0.38 (2σ) $R_\oplus$ around Sun-like stars. We examine the limit in rocky planet radii predicted by hydrodynamic escape across a wide range of possible model inputs using 10,000 parameter combinations drawn randomly from plausible parameter ranges. We find a cutoff between rocky and gas-enveloped planets that agrees with the observed cutoff. The large cross-section available for XUV absorption in the extremely distended primitive atmospheres of low mass planets results in complete loss of atmospheres during the ~100 Myr phase of stellar XUV saturation. In contrast, more massive planets have less distended atmospheres and less escape, and so retain thick atmospheres through XUV saturation and then indefinitely as the XUV and escape fluxes drop over time. The agreement between our model and exoplanet data leads us to conclude that hydrodynamic escape plausibly explains the observed upper limit on rocky planet size and few planets (a "valley" or "radius gap") in the 1.5-2 $R_\oplus$ range.

## 1. INTRODUCTION

In the past decade, thousands of exoplanet candidates and diverse planetary systems have been found (e.g. Hatzes 2016). The variety of characteristics observed among these planets has raised many questions about planetary formation and evolution. Of particular interest is how low mass (defined here as less than ~10 $M_\oplus$), rocky planets form and evolve given their potential to support habitable conditions (e.g. Forget & Leconte 2014). Central to this question is whether a rocky planet will accrete and retain a massive $H_2$/He protoatmosphere that represents a non-negligible fraction of the total planetary mass, which may produce uninhabitable surface temperatures, or if such an atmosphere will be lost. These $H_2$/He protoatmospheres can form on even the smallest planets because once a protoplanet reaches ~0.1 $M_\oplus$ it can accrete $H_2$/He directly from the stellar disk (Hayashi et al. 1979; Ikoma & Hori 2012; Lammer et al. 2011).

If low mass planets form after the protoplanetary nebula dissipates they will not accrete $H_2$/He protoatmospheres (e.g. Massol et al. 2016). However, from planetary formation models, 1-10 $M_\oplus$ planets with periods <100 days may have formed with longer periods when the stellar disk was still present and migrated to their observed locations through interactions with the disk (e.g. Cossou et al. 2013; Raymond & Cossou 2014). The planets considered in this study have short periods (less than 100 days) and thus may have formed before the stellar disk dissipated. Theoretical calculations suggest that such planets may form with $H_2$/He protoatmospheres of 1-10 wt. % (Bodenheimer & Lissauer 2014; Ikoma & Hori 2012; Inamdar & Schlichting 2015). Indeed, a number of low mass exoplanets with thick atmospheres have been observed indicating there is likely no barrier for such protoatmospheres to form on low mass planets (e.g. Cubillos et al. 2016; Lissauer et al. 2013; Masuda 2014).

Observations and subsequent analysis of have shown that rocky planets, without thick protoatmospheres, are only found up to 1.5-2 $R_\oplus$ in size (Dressing et al. 2015; Marcy et al. 2014; Rogers 2015; Weiss & Marcy 2014) with some recent observations indicating that rocky planets can reach ~1.9 $R_\oplus$ in size (Buchhave et al. 2016; Demory et al. 2016). If most close-orbiting planets indeed form with thick protoatmospheres then the least massive planets must have lost their captured $H_2$ and He (see Lopez and Rice (2016) for a discussion of atmospheric formation vs. atmospheric loss). Thermally driven atmospheric loss in the hydrodynamic escape regime is thought to be able to effectively strip the protoatmospheres from planets up to 5-10 $M_\oplus$ (e.g. Chen & Rogers 2016; Howe & Burrows 2015; Jin et al. 2014; Lopez & Fortney 2013; Lopez et al. 2012; Lopez & Rice 2016; Owen & Jackson 2012; Owen & Morton 2016; Owen & Wu 2013, 2016; Wolfgang & Lopez 2015).

The rapid loss of the protoatmosphere via hydrodynamic escape is driven by the XUV emissions (where XUV is X-ray plus extreme UV) from the host star, which heats the upper atmosphere of the planet. For young, Sun-like stars, this XUV flux can be orders of magnitude larger than the modern Sun (Johnstone et al. 2015; Lammer et al. 2014). A saturated XUV flux can last for ~100 Myr (Jackson et al. 2012; Lammer et al. 2012; Ribas et al. 2005). The XUV-driven hydrodynamic escape of a protoatmosphere will occur largely during this saturation time after which the XUV flux and XUV driven hydrodynamic escape decrease exponentially. While Sun-like stars can erode atmospheres from closely orbiting planets (0.1 AU for this model), the modern Earth is orbiting at a sufficient distance that its hydrogen-poor atmosphere is not subjected to hydrodynamic escape (see Catling & Kasting (2017) p. 175 for a discussion of the topic). However, evaporation of Earth-like planets may occur at orbital distances similar to the modern Earth for some low mass planets via water vapor photolysis and subsequent hydrogen escape (e.g. Kasting et al. 2015; Luger & Barnes 2015).

Several recent numerical studies on XUV-driven hydrodynamic escape from low mass planets have results that overlap but differ from each another when simulating the loss of protoatmospheres. Lopez and Fortney (2013), using the model of Lopez et al. (2012), showed that planets less than ~4 $R_\oplus$ could easily lose their atmospheres from hydrodynamic escape, and follow-on work by Lopez and Fortney (2014) suggested that 1.75 $R_\oplus$ was a likely upper size limit for rocky bodies. Similarly, Wolfgang and Lopez (2015) applied the model from Lopez et al. (2012) to data from the *Kepler* mission and found planets with radii above 2 $R_\oplus$ should have atmospheres of at least ~1 wt.% while planets below 2 $R_\oplus$ should have atmospheres less than 1 wt.%. A parameter study of Howe and Burrows (2015) for XUV-driven hydrodynamic escape found that a cutoff between rocky and gas-enveloped planets occurs between $2\,M_\oplus$ and $10\,M_\oplus$ depending on the model orbital distance, which corresponds to a cutoff of $1.26\,R_\oplus$ to $2.16\,R_\oplus$ for an Earth-like density of 5.5 g cm$^{-3}$.

Other studies have discussed an apparent "valley" or "radius gap" in the distribution of exoplanet sizes. Owen and Wu (2013) showed that XUV-driven hydrodynamic escape from low mass planets results in rocky planets less than $1.5\,R_\oplus$, gas-enveloped planets above $2.5\,R_\oplus$, and a lack of planets with intermediate radii. Jin et al. (2014) studied several hydrodynamic escape models and found a valley between 1-2.5 $R_\oplus$ depending on the orbital distance chosen for their model, below which planets are rocky and above which planets have thick atmospheres, typically of at least a few wt. %. Similarly, the XUV-driven hydrodynamic escape model of Chen and Rogers (2016) found a valley in the range of 1-2 $R_\oplus$.

In this study, using an XUV-driven hydrodynamic escape model in which only the atmospheric mass changes over time, we look at the likelihood that XUV-driven hydrodynamic escape can reproduce the observed $1.62^{+0.67}_{-0.08}\,R_\oplus$ cutoff from Rogers (2015) and seek to examine the dominant factors that lie behind the cutoff. We do so by running our model with parameter ranges that describe the most escape-vulnerable planets studied by Rogers (2015). From these escape-vulnerable planets, our model provides an upper limit on the atmospheric loss rate and thus the radii of planets that can lose their entire protoatmospheres and become rocky.

We calculate the cutoff between rocky and gas-enveloped planets with 10,000 different model parameter combinations. We consider only Sun-like stars in this work because the planets used in the study of Rogers (2015) all orbited stars with effective temperatures between 4700 K and 6300 K. The atmospheric loss from XUV-driven hydrodynamic escape around M dwarfs has been considered elsewhere (e.g. Luger et al. 2015; Tian 2009) and we do not address such systems here.

## 2. METHODS

During hydrodynamic escape, a high altitude portion of an atmosphere is heated by XUV flux and flows hydrodynamically outward (Johnstone et al. 2015; Mordasini et al. 2012). However, for thick protoatmospheres that likely represent at least a few wt.% of a planet (Bodenheimer & Lissauer 2014; Ikoma & Hori 2012; Inamdar & Schlichting 2015), the bulk of the lower atmosphere will approximately remain in hydrostatic equilibrium. As such, we assume for this model that above the XUV absorption level, $R_{XUV}$, where the optical depth for the XUV is near unity, the atmosphere is in the hydrodynamic regime and below $R_{XUV}$ the atmosphere is in approximate hydrostatic equilibrium.

The energy-limited rate of XUV-driven hydrodynamic escape from a planet can be approximated by a first-order equation, as follows

$$\frac{dM}{dt} = \frac{\eta \pi F_{XUV} R_{XUV}^3}{GM_p} \tag{1}$$

where $dM/dt$ is the rate of hydrodynamic escape in kg s$^{-1}$ (Watson et al. 1981). The parameter $\eta$ in equation (1) is an efficiency factor that is typically taken to be $0.1 < \eta < 0.6$ (e.g. Bolmont et al. 2017; Koskinen et al. 2014; Lammer et al. 2013; Owen & Wu 2013). The XUV flux incident on the planet in W m$^{-2}$ is given by $F_{XUV}$, $R_{XUV}$ is the radial distance from the planetary center at which the optical depth for broadband XUV radiation is unity, $G$ is the gravitational constant, and $M_p$ is the mass of the planet. For the range of $F_{XUV}$ values considered in this study, the rate of hydrodynamic escape may border on the recombination-limited regime described by Murray-Clay et al. (2009). In the recombination-limited regime, for large XUV fluxes (greater than ~10 W m$^{-2}$), the protoatmospheres could lose energy via recombination of ionized gas slowing the hydrodynamic loss rate. In this recombination-limited regime $dM/dt \propto F_{XUV}^{1/2}$, while $dM/dt \propto F_{XUV}$ in the energy-limited regime. However, on small planets the rate of hydrodynamic escape is dominated by the $R_{XUV}$ term in equation (1) so the difference between the recombination-limited and energy-limited regimes is likely small. As such, we consider only the energy-limited case in this study. To find $dM/dt$ we need only determine the XUV flux and $R_{XUV}$.

The XUV flux from young FGK stars is largest for the first 100±20 Myr after formation (Jackson et al. 2012; Lammer et al. 2012; Ribas et al. 2005). Emissions of XUV are saturated during that time and remain approximately constant. Afterwards, the XUV flux diminishes exponentially and the hydrodynamic loss rate of a planetary atmosphere drops with it. During the saturated regime, for a Sun-like star, the XUV flux can reach ~0.1% of the bolometric luminosity (Jackson et al. 2012; Lammer et al. 2014). Given the uncertainty of stellar evolution a Sun-like star could generate 43-172 W m$^{-2}$ in the XUV at 0.1 AU following Pizzolato et al. (2003). For comparison, at 1 AU the present Earth receives an XUV flux of only ~5 mW m$^{-2}$ (Lammer et al. 2014).

Around Sun-like stars, the protoatmospheres of some low mass planets orbiting interior to 0.1 AU will likely extend beyond the planet's Roche lobe and be rapidly lost (Ginzburg & Sari 2017; Owen & Wu 2013). Not only would the atmosphere be rapidly removed from the high flux and gravitational effects of the host star interior to 0.1 AU, but the rocky core could begin to evaporate as well (Perez-Becker & Chiang 2013). The planets used in the Rogers (2015) work fall mostly outside this limit with only four planets, Kepler-10b, Kepler-21b, Kepler-98b, and Kepler-407b, receiving a flux greater than the 0.1 AU equivalent around a Sun-like star. Kepler-10b, Kepler-21b, and Kepler-407b are less than ~1.6 $R_\oplus$ in radii. In contrast, Kepler-98b has a radius of $1.99\pm0.22 R_\oplus$, with a density of $2.18\pm1.21$ g cm$^{-3}$ so it is not a rocky planet (Marcy et al. 2014). Of the three rocky planets none exceed the radius limit described by Rogers (2015). Thus, we use an orbital distance of 0.1 AU as the inner bound for rocky planets in our model.

We assume that absorption of XUV occurs downward through an upper atmosphere and is fully absorbed by the base of a thermosphere. The problem of finding $R_{XUV}$ then becomes a matter of finding the radial distance to the base of the thermosphere. For a neutral H$_2$ atmospheric column, the broadband XUV flux is typically absorbed within a column of density $10^{26}$ m$^{-2}$ (e.g. Cecchi-Pestellini et al. 2006; Ercolano et al. 2009; Glassgold et al. 2004; Owen & Jackson 2012). For an Earth mass planet, gravity assumed constant, this corresponds to a pressure at the base of the thermosphere of $p_{XUV}=3.3$ Pa and $p_{XUV}=7.1$ Pa for a planet of 10 $M_\oplus$. However, on the modern Earth the base of the thermosphere can occur at pressures as low as $p_{XUV}=0.1$ Pa (Catling & Kasting 2017, p. 4) so we will consider a range of pressures from $0.1 \leq p_{XUV} \leq 10$ Pa. Once $p_{XUV}$ is known, it remains fairly constant over a planet's lifetime even if the surface pressure changes by orders of magnitude (Erkaev et al. 2013). Thus, as rocky planets lose their substantial protoatmospheres, $p_{XUV}$ does not change but rather moves closer to the planetary surface as surface pressure drops.

The protoatmosphere of a young planet will be in approximate hydrostatic equilibrium from the surface to the base of the thermosphere. The radial distance to the base of the thermosphere, $R_{XUV}$, can then be found via the hydrostatic equation, which can be written as

$$-R_g T \frac{1}{p} dp = g_s \left(\frac{R_s}{R}\right)^2 dR \qquad (2)$$

for specific gas constant $R_g$, pressure $p$, isothermal temperature $T$, surface radius of the rocky core $R_s$, and surface gravity $g_s$. In equation (2) we have approximated the gravity term of the hydrostatic equation as $g(R) = g_s [R_s / R]^2$ (see Catling and Kasting (2017), p. 11 for a discussion of the topic). In this study, we assume all rocky planets form with an Earth-like structure and relate mass to radius via

$$R_s = 1.3 M_p^{0.27} \qquad (3)$$

which provides a good approximation for the rocky cores of planets in our model. Equation (3) is derived from the relation $R_s \propto M_p^{0.27}$ given by Zeng et al. (2016). To make this relationship hold when using SI units for the mass ($5.9742 \times 10^{24}$ kg) and radius ($6.371 \times 10^6$ m) of the Earth a scalar value of 1.3 must be used, as seen in equation (3). Integrating equation (2) from the surface to $p_{XUV}$ and solving for $R$ we find

$$R_{XUV} = \frac{R_s^2}{H \ln(p_{XUV} / p_s) + R_s} \qquad (4)$$

with scale height $H = R_g T / g_s$. Surface pressure, $p_s$, can be defined as

$$p_s(M) = \frac{g_s (\alpha M_p - M)}{4 \pi R_s^2} \qquad (A5)$$

where $\alpha$ is the initial protoatmospheric mass fraction, and $M$ is the integrated mass loss from equation (1). We assume an initial $\alpha$ for our model, leaving only the scale height $H$ unknown.

The atmospheric scale height depends on the temperature profile and atmospheric composition. We are interested in the upper limit on rocky planet radii so, as an upper limit, the atmospheric composition was assumed to be pure $H_2$ with $R_g = 4157$ J kg$^{-1}$ K$^{-1}$. This provides an upper bound on $H$ and thus on the loss rate and the radii limit. The atmospheres of Uranus and Neptune have specific gas constants of $R_g \approx 3600$ J kg$^{-1}$ K$^{-1}$ (Lunine 1993) and may represent a composition similar to the protoatmospheres we are modeling so we consider the range $3600 \leq R_g \leq 4157$ J kg$^{-1}$ K$^{-1}$. To calculate the scale height with $R_g$ we use an isothermal atmospheric temperature.

For the protoatmospheres in this study, that represent up to a few wt. % of the total planetary mass, a reasonable upper limit on surface temperature, and thus the isothermal upper atmospheric temperature, is ~3000 K based on temperature profiles from more complex calculations (Jin et al. 2014; Mordasini et al. 2012). In addition, on larger planets with H$_2$ dominated atmospheres, cooling from gas expansion and Lyman-alpha radiation in the upper atmosphere likely result in temperatures of ~3000 K in the thermosphere (Murray-Clay et al. 2009). We expect similar processes to occur in the protoatmospheres of low mass planets. With 3000 K being a reasonable upper limit at both the planetary surface and the base of the thermosphere, we set the upper limit for the isothermal atmospheric temperature in our model to 3000 K. For a lower bound we set the isothermal atmospheric temperature equal to the effective temperature at 0.1 AU. For a Bond albedo of 0 this distance corresponds to an effective temperature of 880 K around a Sun-like star.

The atmospheric loss rate is easily calculated from equation (1). First, an orbital distance is chosen (0.1 AU in this model) and a planetary mass. To calculate the atmospheric loss rate and thus the cutoff between rocky and gas-enveloped planets, 7 additional model parameters must be specified. These parameters are: isothermal atmospheric temperature, $T$; XUV flux at the given orbital distance, $F_{XUV}$; escape efficiency, $\eta$; initial atmospheric mass fraction, $\alpha$; pressure at the base of the thermosphere, $p_{XUV}$; specific gas constant of the atmosphere, $R_g$; and XUV saturation time, $\tau$. We consider 10,000 combinations of these 7 parameters which were selected randomly from a uniform distribution from the values in Table 1. The result for the radius cut-off and its uncertainty is insensitive to the exact number of parameter combinations, whether 10,000 or 20,000 or 5,000 based on sensitivity tests. For each parameter combination, we calculate the cutoff between rocky and gas-enveloped planets using a time step of 10,000 years, which runs quickly and is as accurate as smaller step sizes. Figure 1A and Figure 1B show an example of a model run where the cutoff occurs in the 1.2-1.6 $R_\oplus$ range for the given parameters. Supplemental material ([lehmer.us/#orlpl17](lehmer.us/#orlpl17)) provides an animated version of the model shown in Figure 1B where one can easily see the how gas-enveloped rocky cores of low mass planets evolve into dense rocky planets that become bereft of their primordial atmospheres over time.

**Table 1.** Model Parameter Ranges[†]

| Parameter | Range | Units |
|---|---|---|
| Isothermal atmospheric temperature | $880 \leq T \leq 3000$ | K |
| XUV Flux | $43 \leq F_{XUV} \leq 172$ | W m$^{-2}$ |
| Initial atmospheric mass fraction | $0.01 \leq \alpha \leq 0.1$ | Dimensionless |
| Escape efficiency | $0.1 \leq \eta \leq 0.6$ | Dimensionless |
| Pressure at the base of the thermosphere | $0.1 \leq p_{XUV} \leq 10$ | Pa |
| Specific gas constant | $3600 \leq R_g \leq 4157$ | J kg$^{-1}$ K$^{-1}$ |
| XUV saturation time | $80 \leq \tau \leq 120$ | Myr |

[†]The range of values considered for each parameter used to calculate the cutoff between rocky and gas-enveloped worlds for planets orbiting at 0.1 AU around a Sun-like star. The ranges in the table represent a reasonable upper limit for each parameter from the literature. Justification for each range is given in the text.

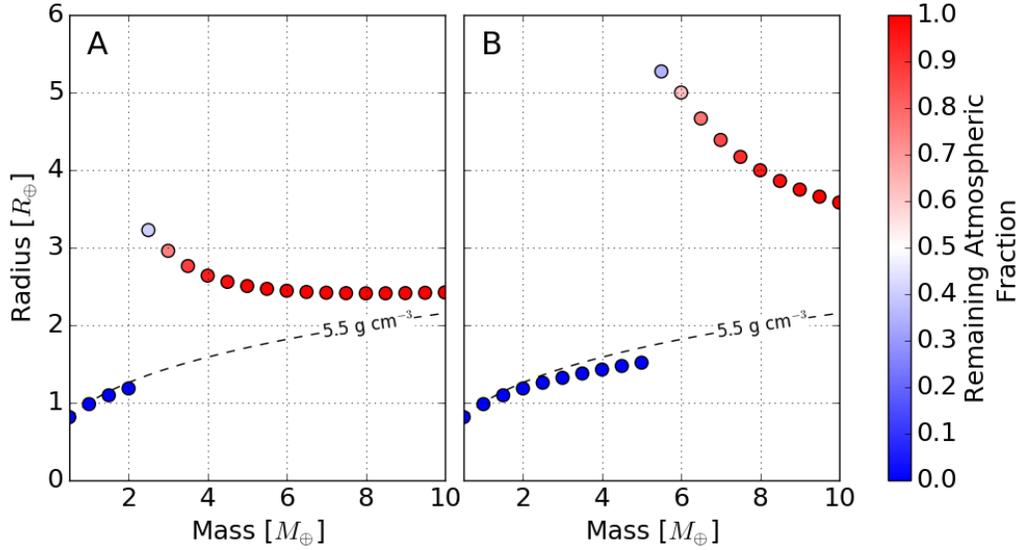

**Figure 1.** Atmospheric loss from planets between 0.5 and 10 Earth masses over 100 Myr with an initial H$_2$ atmosphere of 3 wt.% at 0.1 AU around a young, Sun-like star. The dashed curve shows the contour of fixed Earth-like density of 5.5 g cm$^{-3}$. Blue dots representing rocky bodies fall below this line due to compression at high mass. In both plots the model was run with $\tau = 100$ Myr, $R_g = 4157$ J kg$^{-1}$ K$^{-1}$, $p_{XUV} = 5$ Pa, $F_{XUV} = 55$ W m$^{-2}$, $\eta = 0.1$, and $\alpha = 0.03$. The planets in plot A had an isothermal atmospheric temperature of $T = 880$ K (corresponding to the effective temperature at 0.1 AU with a Bond albedo of 0), and the temperature was set to $T = 1760$ K in plot B. In both cases, we see a sharp cutoff between rocky and gas-enveloped planets occurring in the 1.2 $R_\oplus$ to 1.6 $R_\oplus$ range.

## 3. RESULTS

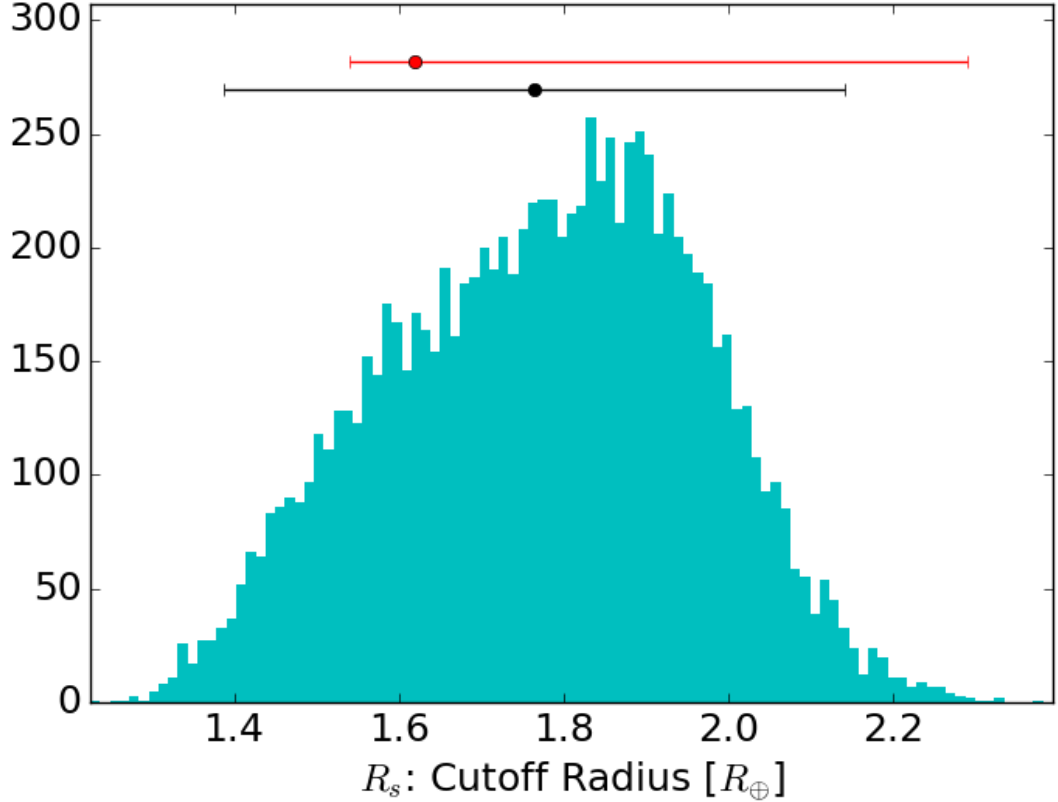

**Figure 2.** 10,000 random parameter combinations were considered using a uniform distribution of the ranges given in Table 1 to calculate the cutoff between rocky and gas-enveloped planets. The number of resulting surface radii, $R_s$, at which the cutoff occurred for each parameter combination is shown in the histogram. The black dot and error bar show the mean cutoff of the distribution at $1.76\pm0.38\,R_\oplus$ from our model with $2s$ uncertainty. The red dot shows the observed $1.62^{+0.67}_{-0.08}\,R_\oplus$ measurement with $2s$ uncertainty from Rogers (2015).

We examined the protoatmospheric loss from planets between 0.5 and $10\,M_\oplus$. The results can be seen in Figure 2, which shows the counts of the calculated cutoff radii, $R_s$, between rocky and gas-enveloped planets for the 10,000 random parameter combinations. The red dot shows the observed cutoff radius for rocky planets from Rogers (2015) with a $2s$ uncertainty. The black dot and error bar shows the model mean and $2s$ uncertainty. The distribution in Figure 2 aligns well with the observed rocky planet limit with both the mean and mode falling within the 95% confidence interval of the observed rocky planet limit. While the mean of our model is 1.76±0.38 (2σ) $R_\oplus$, the mode falls closer to ~1.9 $R_\oplus$, which is where the largest rocky planets have been found. Our model predicts that beyond ~1.9 $R_\oplus$, there is a fairly sharp drop off in the likelihood that hydrodynamic escape can erode a planet. This agrees with recent observations that the largest rocky planets are found up to ~1.9 $R_\oplus$ (Buchhave et al. 2016; Demory et al. 2016). The largest cutoffs predicted in our model are due to parameter combinations with high isothermal atmospheric temperatures and large specific gas constants. In addition to planetary mass, these two parameters control the $R_{XUV}$ term in equation (1), which dominates the loss rate, as discussed below.

Upon close examination, we find that a key aspect of the cutoff between gas-enveloped and rocky planets is that the atmosphere remains very distended, up to several planetary radii in size, and available for XUV absorption on low mass bodies even as atmospheric mass is lost. That protoatmospheres remain puffy, even at low mass, is a result of the logarithmic term in equation (4) generating large values for $R_{XUV}$ until the atmosphere is completely removed. This is seen in Figure 3 where we show $R_{XUV}$ of a 2 $M_\oplus$ planet over time. Even when less than 20% of the original protoatmosphere remains at 20.3 Myr, the radius $R_{XUV}$ is roughly twice the radius of the core. The large radius is caused by the lower gravity on low mass planets coupled to high temperatures (up to several thousand Kelvin), and light atmospheric compositions of $H_2$/He that lead to substantial scale heights and $R_{XUV}$ values that increase rapidly with decreasing mass. In equation (1) we see that the $R_{XUV}$ term is cubed, and it is the only term that changes greatly as mass is lost (for a 3 wt.% atmosphere $M_p$ will only change by at most 3%). Thus, any change in the loss rate will be dominated by the $R_{XUV}^3$ term.

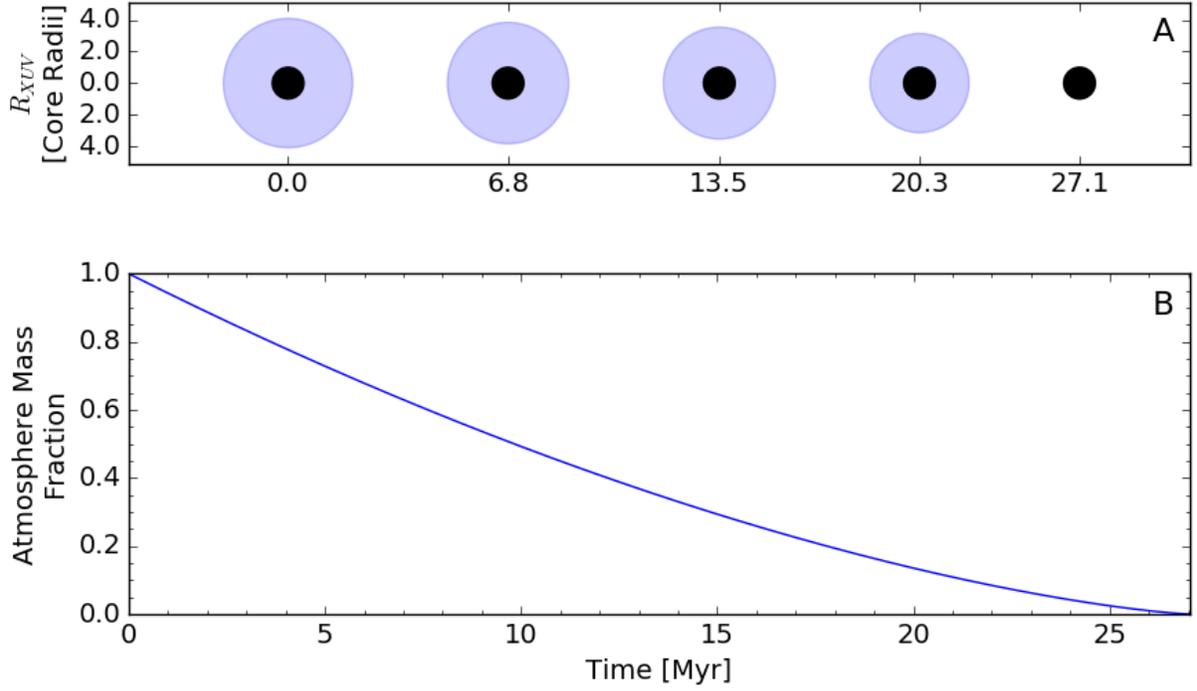

**Figure 3.** A model run for a $2\,M_\oplus$ planet with isothermal atmospheric temperature set to $T = 880$ K. The planet was assumed to orbit at 0.1 AU around a young, Sun-like star with the initial atmosphere representing 3 wt.%. The model parameters were set as $T = 880$ K, $F_{XUV} = 55$ W m$^{-2}$, $\alpha = 0.03$, $\eta = 0.1$, $p_{XUV} = 5$ Pa, $R_g = 4157$ J kg$^{-1}$ K$^{-1}$, and $\tau = 100$ Myr. Plot A shows a snapshot of the planetary radius up to $R_{XUV}$ at times of 0, 6.8, 13.5, 20.3, and 27.1 Myr in the simulation. The blue region shows the relative size of the planetary atmosphere, and the black region shows the size of the rocky core. Plot B shows the remaining atmospheric mass fraction over time. We see that $R_{XUV}$ remains large even after most of the atmosphere has been lost.

In Figure 4, we see how, for planets with mass $< 2.5\,M_\oplus$, $R_{XUV}^3$ is orders of magnitude larger than $R_\oplus$ while the protoatmosphere remains, but for planets with mass 6-7 $M_\oplus$, $R_{XUV}^3$ levels off to ~17 $R_\oplus$. Not only do low mass planets have much larger loss rates due to this exponential increase in $R_{XUV}$, but they also have less overall atmospheric mass to lose. The strong nonlinearity of the hydrostatic equation shows us that there will exist a critical planet size below which $R_{XUV}$ increases rapidly leading to substantial hydrodynamic escape. This non-linear dependence on $R_{XUV}$ has been noted in previous work (e.g. Chen & Rogers 2016; Lopez & Fortney 2013) and, on average in our model, results in planets with cores larger than 1.76±0.38 (2σ) $R_⊕$ retaining a significant portion of their protoatmospheres. The largest planets that can lose their entire protoatmosphere are thus planets with rocky cores less ~1.8 $R_⊕$.

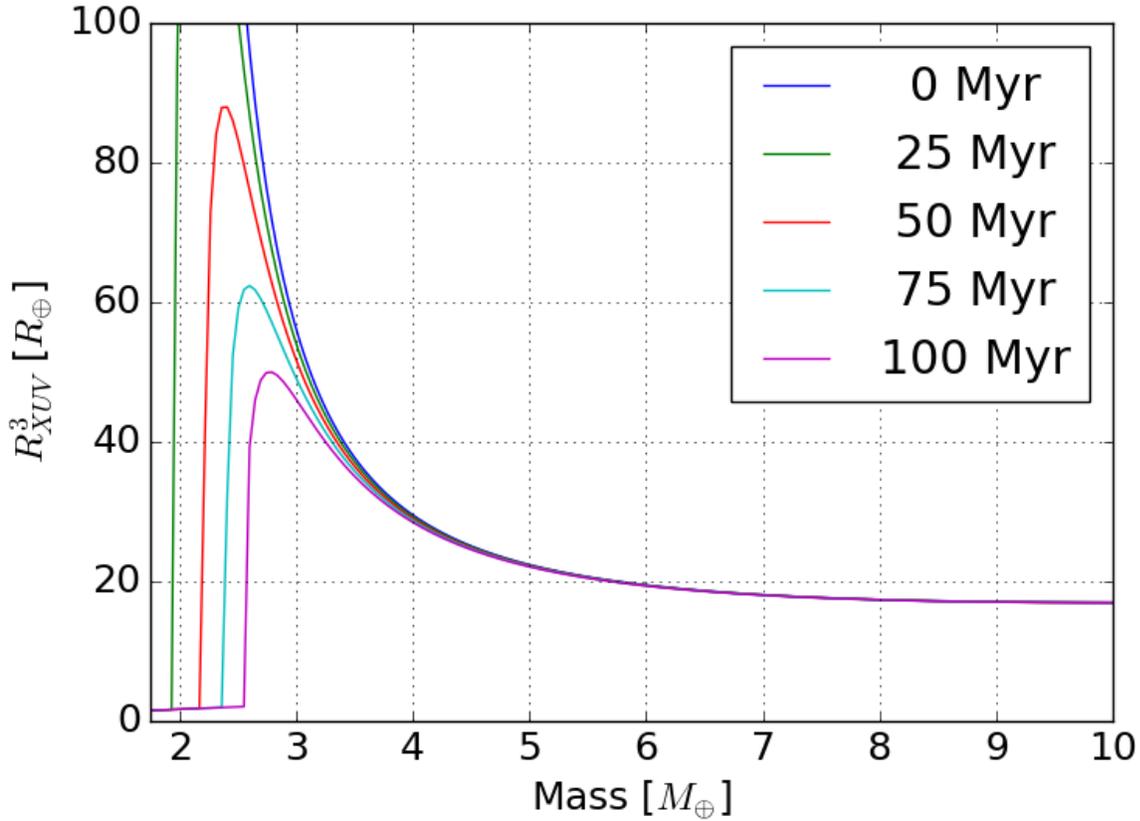

**Figure 4.** The value of $R_{XUV}^3$ at times $\tau = 0$, 25, 50, 75, and 100 Myr (shown by the blue, green, red, cyan, and magenta curves, respectively) for planets between 1.75 $M_⊕$ and 10 $M_⊕$ around a Sun-like star at 0.1 AU. The model parameters were set to $T = 880$ K, $F_{XUV} = 55$ W m$^{-2}$, $\alpha = 0.03$, $\eta = 0.1$, $p_{XUV} = 5$ Pa, and $R_g = 4157$ J kg$^{-1}$ K$^{-1}$. The low mass planets have large $R_{XUV}^3$ values that cause rapid loss. By ~6 $M_⊕$, $R_{XUV}^3$ has become roughly constant with mass. The approximately flat line at low masses (below ~2.5 $M_⊕$ for the magenta curve) indicates that, for a given $\tau$, the atmosphere has been entirely lost and $R_{XUV}^3$ is at the surface of the planet.

In addition to the predicted cutoff at 1.8 $R_\oplus$, our model shows that there should be a lack of planets with radii immediately larger and smaller than the cutoff radius. This is seen in Figure 1 where the abrupt jump from rocky to gas-enveloped planets may result in a void where planets are unlikely to exist. This agrees with a number of previous studies (see Introduction) and the recent work by Fulton et al. (2017) which found that such a deficit is indeed present in the current exoplanet data.

The exoplanet data from Fulton et al. (2017) is shown in Figure 5 with our model predictions. Our model indicates that for radii below ~2 $R_\oplus$ planets are less likely to be gas-enveloped (the blue region in Figure 5), and for radii above ~1.5 $R_\oplus$ planets are less likely to be rocky (the red region in Figure 5). The paucity of 1.5-2 $R_\oplus$ planets predicted by our model is seen in the Fulton et al. (2017) data. The missing planets fall into the evaporation valley, or radius gap described by previous XUV-driven hydrodynamic escape studies (see Introduction) and show that XUV-driven hydrodynamic escape is able to reproduce the major characteristics of the observed, low mass exoplanet population.

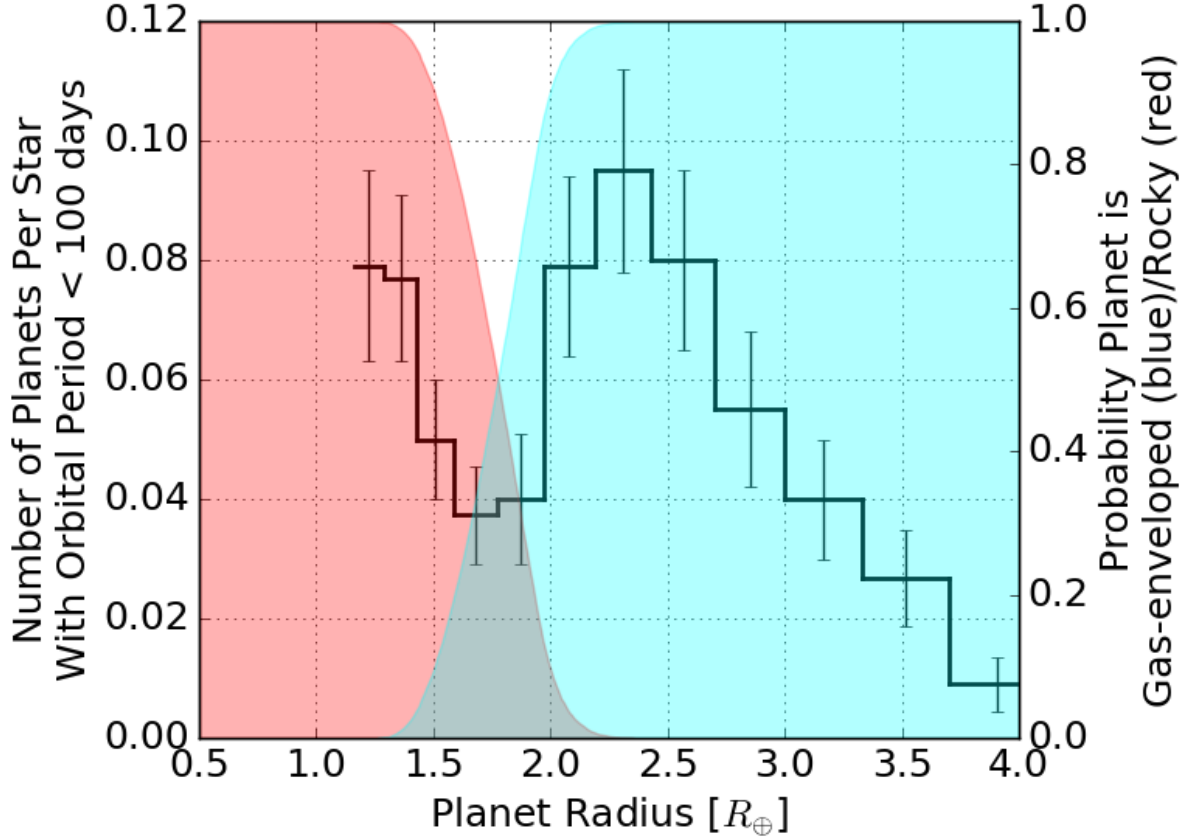

**Figure 5.** The exoplanet data from Fulton et al. (2017) is shown in the black curve (data taken from Fulton et al. (2017) Table 3). Our model predictions are shown by the shaded regions. Using the probability distribution generated in Figure 2 the red shaded region shows the probability that a planet is below the rocky planet cutoff, and the blue shaded region shows the probability that a planet is above the rocky planet cutoff. Beyond ~1.5 $R_\oplus$ planets are unlikely to be rocky while below ~2 $R_\oplus$ planets are unlikely to be gas-enveloped. Thus, our model predicts a lack of exoplanets with radii between 1.5-2 $R_\oplus$ which is indeed seen in the Fulton et al. (2017) data. Fulton et al. (2017) found that planets below ~1.8 $R_\oplus$ are likely to be rocky while larger planets are likely to be gas-enveloped, with which our results agree. It should be noted that the vertical axes in this plot are arbitrarily scaled. The important feature, however, is the location of the transition between rocky and gas-enveloped planets and the width of the valley between that transition, which is agnostic of the vertical scaling.

## 4. DISCUSSION

The transition from rocky to gas-enveloped planets occurs where XUV-driven hydrodynamic escape predicts such a transition should occur (see Figure 5). That the transition has been predicted across numerous studies with models of varying complexity (see Introduction) and agrees with current exoplanet data leads us to conclude that hydrodynamic escape is plausibly the cause of the observed limit in rocky planet radii. The closely orbiting exoplanets (periods less than ~100 days) modeled in this study comprise the majority of known, low mass exoplanets (e.g. Batalha 2014). That XUV-driven hydrodynamic atmospheric escape is important for these planets is not surprising given the large XUV fluxes present at such short orbital periods.

As additional, longer period rocky planets are discovered the average cutoff in rocky planet size may decrease with increasing orbital period, as noted by Lopez and Rice (2016). However, at large orbital distances where the XUV flux is small and XUV-driven hydrodynamic escape becomes negligible other processes may limit the size of rocky planets. Indeed, Zeng et al. (2017) found a bimodal distribution in the current exoplanet data similar to Fulton et al. (2017) but note that it could be explained by formation scenarios rather than evolutionary ones (i.e. XUV-driven hydrodynamic escape). The radius limit for closely orbiting rocky planets appears to be set at ~1.8 $R_\oplus$ by XUV-driven hydrodynamic escape, but to address the limit in rocky planet size for longer period planets, additional studies on rocky planet formation should be conducted.

**ACKNOWLEDGEMENTS**
We would like to thank Kevin Zahnle and Joshua Krissansen-Totton for their advice and guidance on this paper. D.C.C., was supported by NASA Planetary Atmospheres grant NNX14AJ45G. O.R.L. was supported by the NASA Pathways Program.

References

Batalha, N. M. 2014, Proc Natl Acad Sci, 111, 12647
Bodenheimer, P., & Lissauer, J. J. 2014, Astrophys J, 791, 103
Bolmont, E., Selsis, F., Owen, J. E., et al. 2017, Mon Not R Astron Soc, 464, 3728
Buchhave, L. A., Dressing, C. D., Dumusque, X., et al. 2016, Astron J, 152, 160
Catling, D. C., & Kasting, J. F. 2017, Atmospheric Evolution on Inhabited and Lifeless Worlds (New York: Cambridge University Press)
Cecchi-Pestellini, C., Ciaravella, A., & Micela, G. 2006, Astron Astrophys, 458, L13
Chen, H., & Rogers, L. A. 2016, Astrophys J, 831, 180
Cossou, C., Raymond, S. N., & Pierens, A. 2013, Proc Int Astron Union, 8, 360
Cubillos, P., Erkaev, N. V., Juvan, I., et al. 2016, Mon Not R Astron Soc, stw3103
Demory, B.-O., Gillon, M., De Wit, J., et al. 2016, Nature, 532, 207
Dressing, C. D., Charbonneau, D., Dumusque, X., et al. 2015, Astrophys J, 800, 135
Ercolano, B., Clarke, C. J., & Drake, J. J. 2009, Astrophys J, 699, 1639
Erkaev, N. V., Lammer, H., Odert, P., et al. 2013, Astrobiology, 13, 1011
Forget, F., & Leconte, J. 2014, Philos Trans R Soc Math Phys Eng Sci, 372, 20130084
Fulton, B. J., Petigura, E. A., Howard, A. W., et al. 2017, ArXiv170310375 Astro-Ph, http://arxiv.org/abs/1703.10375
Ginzburg, S., & Sari, R. 'em. 2017, Mon Not R Astron Soc, 464, 3937
Glassgold, A. E., Najita, J., & Igea, J. 2004, Astrophys J, 615, 972
Hatzes, A. P. 2016, Space Sci Rev, 205, 267
Hayashi, C., Nakazawa, K., & Mizuno, H. 1979, Earth Planet Sci Lett, 43, 22
Howe, A. R., & Burrows, A. 2015, Astrophys J, 808, 150
Ikoma, M., & Hori, Y. 2012, Astrophys J, 753, 66
Inamdar, N. K., & Schlichting, H. E. 2015, Mon Not R Astron Soc, 448, 1751
Jackson, A. P., Davis, T. A., & Wheatley, P. J. 2012, Mon Not R Astron Soc, 422, 2024
Jin, S., Mordasini, C., Parmentier, V., et al. 2014, Astrophys J, 795, 65
Johnstone, C. P., Güdel, M., Stökl, A., et al. 2015, Astrophys J Lett, 815, L12
Kasting, J. F., Chen, H., & Kopparapu, R. K. 2015, Astrophys J Lett, 813, L3
Koskinen, T. T., Lavvas, P., Harris, M. J., & Yelle, R. V. 2014, Phil Trans R Soc A, 372, 20130089
Lammer, H., Erkaev, N. V., Odert, P., et al. 2013, Mon Not R Astron Soc, 430, 1247
Lammer, H., Stökl, A., Erkaev, N. V., et al. 2014, Mon Not R Astron Soc, 439, 3225
Lammer, H., Güdel, M., Kulikov, Y., et al. 2012, Earth Planets Space, 64, 179
Lammer, H., Kislyakova, K. G., Odert, P., et al. 2011, Orig Life Evol Biospheres, 41, 503
Lissauer, J. J., Jontof-Hutter, D., Rowe, J. F., et al. 2013, Astrophys J, 770, 131
Lopez, E. D., & Fortney, J. J. 2013, Astrophys J, 776, 2
Lopez, E. D., & Fortney, J. J. 2014, Astrophys J, 792, 1
Lopez, E. D., Fortney, J. J., & Miller, N. 2012, Astrophys J, 761, 59
Lopez, E. D., & Rice, K. 2016, ArXiv Prepr ArXiv161009390, https://arxiv.org/abs/1610.09390
Luger, R., Barnes, R., Lopez, E., et al. 2015, Astrobiology, 15, 57
Luger, R., & Barnes, R. 2015, Astrobiology, 15, 119
Lunine, J. I. 1993, Annu Rev Astron Astrophys, 31, 217
Marcy, G. W., Isaacson, H., Howard, A. W., et al. 2014, Astrophys J Suppl Ser, 210, 20
Massol, H., Hamano, K., Tian, F., et al. 2016, Space Sci Rev, 205, 153
Masuda, K. 2014, Astrophys J, 783, 53

Mordasini, C., Alibert, Y., Georgy, C., et al. 2012, Astron Astrophys, 547, A112
Murray-Clay, R. A., Chiang, E. I., & Murray, N. 2009, Astrophys J, 693, 23
Owen, J. E., & Jackson, A. P. 2012, Mon Not R Astron Soc, 425, 2931
Owen, J. E., & Morton, T. D. 2016, Astrophys J, 819, L10
Owen, J. E., & Wu, Y. 2013, Astrophys J, 775, 105
Owen, J. E., & Wu, Y. 2016, Astrophys J, 817, 107
Perez-Becker, D., & Chiang, E. 2013, Mon Not R Astron Soc, 433, 2294
Pizzolato, N., Maggio, A., Micela, G., Sciortino, S., & Ventura, P. 2003, Astron Astrophys, 397, 147
Raymond, S. N., & Cossou, C. 2014, Mon Not R Astron Soc Lett, 440, L11
Ribas, I., Guinan, E. F., Güdel, M., & Audard, M. 2005, Astrophys J, 622, 680
Rogers, L. A. 2015, Astrophys J, 801, 41
Tian, F. 2009, Astrophys J, 703, 905
Watson, A. J., Donahue, T. M., & Walker, J. C. G. 1981, Icarus, 48, 150
Weiss, L. M., & Marcy, G. W. 2014, Astrophys J Lett, 783, L6
Wolfgang, A., & Lopez, E. 2015, Astrophys J, 806, 183
Zeng, L., Jacobsen, S. B., Hyung, E., et al. 2017, in Terrestrial Planet Differentiation: Everywhere Every Way (The Woodlands, Texas)
Zeng, L., Sasselov, D. D., & Jacobsen, S. B. 2016, Astrophys J, 819, 127